# Enhancing Transportation Cyber-Physical Systems Security: A Shift to Post-Quantum Cryptography

Abdullah Al Mamun, A. A., Mamun*

Ph.D. Student, Glenn Department of Civil Engineering, Clemson University, Clemson, SC 29634, USA, abdullm@clemson.edu

Akid Abrar, A., Abrar

Ph.D. Student, Department of Civil, Construction, and Environmental Engineering, The University of Alabama, Tuscaloosa, AL 35487, USA, aabrar@crimson.ua.edu

Mizanur Rahman, M., Rahman

Assistant Professor, Department of Civil, Construction & Environmental Engineering, The University of Alabama, Tuscaloosa, AL 35487, USA, mizan.rahman@ua.edu

M Sabbir Salek, M. S., Salek

Senior Engineer, National Center for Transportation Cybersecurity and Resiliency (TraCR), Greenville, SC 29607, USA, msalek@clemson.edu

Mashrur Chowdhury, M., Chowdhury

Professor, Glenn Department of Civil Engineering, Clemson University, Clemson, SC 29634, USA, mac@clemson.edu

The rise of quantum computing threatens traditional cryptographic algorithms that secure Transportation Cyber-Physical Systems (TCPS). Shor's algorithm poses a significant threat to RSA and Elliptic Curve Cryptography (ECC), while Grover's algorithm reduces the security of symmetric encryption schemes, such as AES. The objective of this paper is to underscore the urgency of transitioning to post-quantum cryptography (PQC) to mitigate these risks in TCPS by analyzing vulnerabilities of traditional cryptographic schemes and the applicability of standardized PQC schemes in TCPS. We analyzed vulnerabilities in traditional cryptography against quantum attacks and reviewed the applicability of the National Institute of Standards and Technology (NIST)-standardized PQC schemes, including CRYSTALS-Kyber, CRYSTALS-Dilithium, and SPHINCS+, in TCPS. We conducted a case study to analyze the vulnerabilities of a TCPS application from the Architecture Reference for Cooperative and Intelligent Transportation (ARC-IT) service package, i.e., Electronic Toll Collection, leveraging the Microsoft Threat Modeling tool. This case study highlights the cryptographic vulnerabilities of a TCPS application and presents how PQC can effectively counter these threats. Additionally, we evaluated CRYSTALS-Kyber's performance across wired and wireless TCPS data communication scenarios. While CRYSTALS-Kyber proves effective in securing TCPS applications over high-bandwidth, low-latency Ethernet networks, our analysis highlights challenges in meeting the stringent latency requirements of safety-critical wireless applications within TCPS. Future research should focus on developing lightweight PQC solutions and hybrid schemes that integrate traditional and PQC algorithms to enhance compatibility, scalability, and real-time performance, ensuring robust protection against emerging quantum threats in TCPS.

---

\* Corresponding author.

# 1 INTRODUCTION

The rapid evolution of connectivity and automation in transportation systems, characterized by the emergence of transportation cyber-physical systems (TCPS), has significantly raised the importance of secure data exchange. TCPS refers to an integrated system where the physical elements, such as transportation infrastructure and vehicles, communicate with the cyber elements via communication technologies, and interact seamlessly to enhance efficiency, safety, and sustainability of transportation systems [20, 50]. Primarily TCPS encompasses intelligent transportation system (ITS), as it serves as a foundational component, ensuring that cyber and physical elements of transportation systems operate cohesively [77]. These interconnected systems rely heavily on cryptographic security to protect and safeguard sensitive information transmitted between vehicles, transportation infrastructure, and other entities.

According to the United States Department of Transportation's (USDOT) Architecture Reference for Cooperative and Intelligent Transportation (ARC-IT), cryptographic security is essential in various TCPS to ensure data integrity, confidentiality, and authenticity [74]. For instance, electronic toll collection systems require encryption and authentication to safeguard payment information, preventing fraud and ensuring accurate toll calculations. Traffic signal control systems use cryptographic methods to protect communication between signals and central control, preventing malicious manipulation. Emergency response systems depend on secure protocols to protect sensitive patient data and communication between first responders. Commercial vehicle operations use encryption and digital signatures to secure fleet data, driver credentials, and cargo information, preventing theft and unauthorized access. Transit security systems safeguard passenger data and payment information, ensuring privacy and preventing identity theft. Connected vehicles use cryptographic techniques to protect data exchange between vehicles and roadside units, ensuring safety and reliability. Infrastructure-based traffic surveillance and weather information systems use encryption and data integrity measures to protect video footage, traffic data, and weather forecasts, ensuring accuracy and preventing unauthorized access or manipulation.

With the advent of quantum computing, the cryptographic algorithms that form the basis of modern security systems are increasingly at risk. Quantum computers, with their ability to perform calculations at exponentially greater speeds than classical computers, could potentially break the current cryptographic protections. Algorithms, such as Shor's algorithm [68], can efficiently factor large numbers and solve discrete logarithm problems, directly compromising widely used schemes, e.g., Rivest–Shamir–Adleman (RSA) [63], Diffie-Hellman (DH) [21], Digital Signature Algorithm (DSA) [17], and Elliptic Curve Cryptography (ECC) (e.g., Elliptic Curve Digital Signature Algorithm [ECDSA] [33]). Furthermore, Grover's algorithm can drastically reduce the complexity of brute-force attacks on symmetric key algorithms, such as the Advanced Encryption Standard (AES) [12, 28]. To mitigate the risks posed by quantum computing, the transportation industry requires a shift towards post-quantum cryptography (PQC). PQC offers a promising avenue for developing cryptographic algorithms that are resistant to quantum attacks. By exploring the integration of PQC into TCPS, we could ensure the long-term security and resilience of these critical infrastructures.

This paper evaluates the quantum vulnerabilities of traditional cryptographic standards and examines the potential of PQC to enhance the security of TCPS. By examining the strengths and weaknesses of traditional cryptographic schemes and PQC schemes, this research seeks to inform the development of robust security strategies for TCPS. To illustrate real-world implications, the study analyzes electronic toll collection as a use case, focusing on vulnerabilities exposed by quantum computers and demonstrating the resilience provided by PQC mechanisms. Additionally, the paper provides a detailed experimental performance evaluation of CRYSTALS-Kyber, a PQC scheme standardized by the National Institute of Standards and Technology (NIST) in 2024, offering practical insights into its suitability for securing TCPS. The primary contribution of this paper is demonstrating the urgency of transitioning to PQC to mitigate cyber risks of secure data communication in TCPS by analyzing the vulnerabilities of traditional cryptographic schemes and an evaluation of the applicability of standardized PQC schemes in TCPS.

To provide a comprehensive overview, this paper is structured as follows: Section 2 presents the traditional cryptographic schemes and their vulnerabilities to quantum attacks. Section 3 discusses the principles and standards of PQC, details the NIST PQC Standardization Process, explores various families of PQC schemes, and compares the claimed security levels and key sizes of NIST fourth-round finalists, including the three schemes standardized in 2024. Section 4 provides a case study on electronic toll collection, illustrating the practical implications of quantum attacks and the potential impact of PQC countermeasures. Section 5 presents an experimental evaluation of CRYSTALS-Kyber, focusing on its security features, including resistance to known attack vectors and quantum resource requirements, and its performance in various peer-to-peer communication scenarios through detailed simulation setups. Section 6 discusses the challenges ahead, outlining potential areas for further research and development in enhancing cryptographic security in TCPS. Finally, Section 7 details the conclusions, summarizing the findings of the study.

## 2 VULNERABILITIES OF TRADITIONAL CRYPTOGRAPHIC SCHEMES TO QUANTUM COMPUTING-BASED ATTACK

Cryptographic schemes can be classified into several types based on their functions: public key cryptography, symmetric cryptography, digital signatures, and hashing functions [46]. Public key cryptography uses a pair of keys- a public key, which can be shared openly, and a private key, which is kept secret. It is primarily used for secure key exchange and digital signatures, with examples including RSA, DH, and ECC [46]. Symmetric cryptography, such as, the AES, uses a single key for both encryption and decryption processes, making it particularly efficient for encrypting large datasets [46]. However, this secret key needs to be securely communicated between parties before encryption and decryption can occur, often requiring additional mechanisms or protocols for key exchange, such as Key Encapsulation Mechanisms (KEMs) [13]. Digital signatures verify the authenticity and integrity of a message, providing non-repudiation; examples include the ECDSA and the DSA [46]. Hashing functions generate a fixed-size hash value from input data, ensuring data integrity, such as SHA-256 [46].

The Institute of Electrical and Electronics Engineers (IEEE) 1609.2 standard for Wireless Access in Vehicular Environments (WAVE) incorporates a suite of cryptographic algorithms to ensure the integrity, confidentiality, and authenticity of data exchanged within TCPS [87]. These algorithms include the ECDSA for digital signatures [33], AES-128 [52] and SM4 [69] for symmetric encryption, and the Elliptic Curve Integrated Encryption Scheme (ECIES) [26] and SM2 [6] for key establishment and encryption. Hash functions such as SHA-256 [30] and SM3 [78] are used for data integrity verification. However, the advent of quantum computing poses significant threats to these cryptographic protections. The next subsections will discuss the vulnerabilities of cryptographic schemes incorporated into the IEEE 1609.2 standard, as well as some commonly used cryptographic schemes.

## 2.1 Vulnerabilities in Public Key Cryptography and Digital Signatures

ECC schemes, including ECDSA and SM2, offer enhanced security compared to traditional methods using smaller key sizes. However, these systems are susceptible to Shor's algorithm [68], a quantum computing algorithm capable of efficiently solving the discrete logarithm problem, which forms the basis of ECC schemes. Shor's algorithm presents a considerable risk to digital signatures and key exchange protocols based on ECC. Research by Kirsch (2015) has extended their algorithm's capabilities to decrypt data encrypted with ECC, highlighting the vulnerability of ECC's smaller key space to quantum attacks [38]. Proos and Zalka's (2003) work further emphasize the potential impact of quantum computing on ECC. Their findings indicate that a 1000-qubit quantum computer could break 160-bit elliptic curves while factorizing a 1024-bit RSA key would require a more powerful 2000-qubit system [59].

RSA relies on the difficulty of factoring large integers [63], a problem that quantum computers can solve efficiently using Shor's algorithm. This makes RSA encryption insecure against quantum attacks. The study by Shakib et al. (2023) demonstrated that a quantum impersonation attack using Shor's algorithm could break RSA-encrypted digital signatures in a blockchain-based Vehicular Ad-hoc Network (VANET) [65]

In addition, both DH and DSA are vulnerable to Shor's algorithm due to their reliance on the discrete logarithm problem. Quantum computers can solve this problem efficiently, compromising the security of these key exchange and digital signature schemes [44]. The implications of this vulnerability are significant, as DH and DSA are widely used in today's cryptographic infrastructure.

## 2.2 Vulnerabilities in Symmetric Key Cryptography

Both AES and SM4 are symmetric encryption algorithms susceptible to the quantum threat posed by Grover's algorithm [28]. While AES generally offers stronger resistance due to its flexibility in key size, even AES-128, with a 128-bit key, can be effectively reduced to a 64-bit key through Grover's algorithm. SM4, using a fixed 128-bit key, faces similar vulnerabilities. Employing larger key sizes, such as AES-256, is necessary to mitigate these quantum risks. The impact of quantum attacks on symmetric encryption is exemplified by the Data Encryption Standard (DES). Bone and Castro (1997) demonstrated that Grover's algorithm could break DES, which uses a 56-bit key, with only 185 search attempts [12]. This underscores the necessity for robust countermeasures against quantum threats in symmetric cryptography.

## 2.3 Vulnerabilities in Hash Functions

Hash functions, which produce fixed-size outputs (hash values) from arbitrary-length inputs [46], are also vulnerable to quantum attacks. Grover's algorithm can significantly accelerate the process of finding collisions in hash functions, reducing their security. Specifically, a hash function with a $2^n$-bit output can be broken with approximately $2^{n/2}$ operations using Grover's algorithm [28]. Furthermore, a technique known as the quantum birthday attack, combining Grover's algorithm with the birthday paradox, can find collisions even more efficiently [41]. To counter these threats, hash functions require substantially longer outputs in the quantum era. For instance, according to Brassard et al. (1997), a hash function aiming for *b*-bits of security against Grover's attacks should produce a *3b*-bit output [14]. While many current hash algorithms fall short of this requirement, SHA-2 and SHA-3, with longer output sizes, offer greater resistance to quantum attacks.

The impact of quantum computing on various traditional cryptographic schemes is further summarized in **Table 1**. This table is adapted from NIST and illustrates how quantum computing compromises the security of traditional cryptographic algorithms, making schemes such as RSA, ECC, and DSA no longer secure [16]. Given the vulnerabilities, it is evident that current cryptographic schemes are insufficient in the face of advancing quantum computing technology. The next section explores PQC as a solution to this critical issue.

Table 1: Impact of Quantum Computing on Traditional Cryptographic Schemes (Adapted from [16])

| Cryptographic Scheme | Type | Purpose | Impact from Large-Scale Quantum Computer |
| --- | --- | --- | --- |
| AES | Symmetric Key | Encryption | Larger key sizes needed |
| SHA-2, SHA-3 | Hash Functions | Hashing | Longer output needed |
| RSA | Public Key | Signatures, Key Establishment | No longer secure |
| ECDSA, ECDH (ECC) | Public Key | Signatures, Key Exchange | No longer secure |
| DSA (Finite Field Cryptography) | Public Key | Signatures, Key Exchange | No longer secure |

## 3 POST-QUANTUM CRYPTOGRAPHY

PQC represents a new class of cryptographic algorithms designed to be secure against attacks from both classical and quantum computers. The primary goals of PQC are to develop cryptographic schemes that are secure against quantum attacks, ensure practical implementation across various applications, and facilitate a smooth transition from classical to quantum-resistant cryptographic systems [16]. Extensive research, public engagement, and the establishment of standards are critical components of this transition. The NIST has been at the forefront of PQC research and standardization. This section aims to provide a comprehensive overview of PQC by discussing the NIST PQC Standardization Process, the various families of PQC schemes, and the security requirements and evaluations that guide the selection of these schemes. We will also compare the claimed security levels, key sizes, and other relevant metrics of the NIST fourth-round finalists to illustrate the diverse approaches and trade-offs in PQC designs.

### 3.1 NIST PQC Standardization Process

To address the growing threat of quantum computers to current encryption standards, NIST initiated a standardization process for PQC in 2016. This multi-round evaluation aimed to select algorithms resistant to quantum attacks. From an initial pool of 69 candidate algorithms for public-key encryption/ KEMs and digital signatures, NIST narrowed down the field to 26 finalists in the second round, and subsequently to 15 in the third [1, 2, 51]. Ultimately, after the third-round evaluation, CRYSTALS-Kyber [13] was chosen for standardization as a public-key encryption/KEM scheme, while CRYSTALS-Dilithium [23], Falcon [80], and Stateless Practical Hash-based Incredibly Nice and Compact Signatures (SPHINCS+) [8] were selected for standardization as digital signatures [2]. On August 13, 2024, NIST finalized the standardization of three schemes: CRYSTALS-Kyber, CRYSTALS-Dilithium, and SPHINCS+, marking a significant milestone in post-quantum cryptography. These schemes have been renamed as part of their formal standardization: CRYSTALS-Kyber is now 'Module-Lattice-Based Key-Encapsulation Mechanism (ML-KEM)' [53], CRYSTALS-Dilithium is 'Module-Lattice-Based Digital Signature Algorithm (ML-DSA)' [54], and SPHINCS+ is 'Stateless Hash-Based Digital Signature Algorithm (SLH-DSA)' [55], reflecting their finalized status in the standards. As the process continues, four additional public-key encryption/KEM candidates (i.e., Bit Flipping Key Encapsulation [BIKE] [81], Classic McEliece [82], Hamming Quasi-Cyclic [HQC] [83], and Supersingular Isogeny Key Encapsulation [SIKE] [84]) advanced to the fourth round [2]. This rigorous process evaluates algorithms based on security, performance, and implementation practicality to ensure robust protection of sensitive data in the quantum era.

## 3.2 PQC Scheme Overview

PQC includes a range of cryptographic algorithms designed to remain secure against attacks from both classical and quantum computers. From the literature review and NIST's extensive evaluation process, five main families of PQC schemes have been identified: lattice-based, code-based, multivariate polynomial-based, hash-based, and isogeny-based. This section briefly discusses the underlying mathematical problems and inherent security properties of these families.

*3.2.1 Lattice-Based Cryptography*

Lattice-based cryptography is built on hard mathematical problems related to lattices, such as the Shortest Vector Problem (SVP) [47] and the Learning with Errors (LWE) [4] problem. These problems involve finding short or close vectors in a high-dimensional lattice, which is computationally infeasible for both classical and quantum computers. The security of lattice-based schemes relies on the assumed hardness of these problems [64]. Furthermore, the structured randomness and noise introduced in the LWE problems create additional layers of difficulty, making it extremely challenging for quantum algorithms like Shor's algorithm or Grover's algorithm to solve efficiently. Reflecting their strong security foundations, the NIST evaluation's third round led to the standardization of one lattice-based public-key encryption/KEM scheme (CRYSTALS-Kyber) and one lattice-based digital signature scheme (CRYSTALS-Dilithium), now formally recognized as 'ML-KEM' and 'ML-DSA', respectively [53, 54]. The standardization process for Falcon, another lattice-based algorithm used as a digital signature scheme, is currently ongoing.

*3.2.2 Code-Based Cryptography*

Code-based cryptography relies on the hardness of problems related to error-correcting codes, particularly the General Decoding Problem (GDP) and the Syndrome Decoding Problem (SDP), both of which are NP-complete [7]. One of the most well-known examples is the McEliece cryptosystem [45], which has been extensively studied and remains resistant to cryptanalytic attacks [9]. No efficient quantum algorithms have been found to solve these decoding problems, making code-based cryptography quantum-resistant. This resilience is why schemes like Classic McEliece have been selected as finalists in the NIST PQC Standardization Process [2]. Classic McEliece offers strong security with fast encryption and decryption, despite its larger key sizes compared to other schemes. Another notable scheme is HQC, which also relies on the hardness of decoding random linear codes [83]. In addition to Classic McEliece, HQC has been selected as a fourth-round finalist in the NIST PQC Standardization Process [2].

*3.2.3 Multivariate Polynomial-Based Cryptography*

Multivariate polynomial-based cryptography relies on the complexity of solving systems of multivariate polynomial equations over finite fields. These problems, such as the Multivariate Quadratic (MQ) problem, are known to be NP-hard, making them computationally infeasible to solve efficiently with either classical or quantum computers [2]. The security of multivariate schemes is based on the assumed difficulty of finding solutions to these polynomial equations. Notable examples of multivariate polynomial-based schemes include Great Multivariate Short Signature (GeMSS) scheme [85] and Rainbow signature scheme [86]. Both schemes offered high security and efficient performance and were initially selected as finalists in the NIST PQC Standardization Process. However, both faced significant cryptanalytic challenges in the third round and therefore, were not selected to progress to the fourth round. GeMSS suffered a key-recovery attack that revealed the private key, making attempts to restore security impractical [70, 71]. Similarly, Rainbow's security was significantly reduced by new attacks, necessitating extensive re-engineering to meet security targets, which negated its performance advantages [2].

*3.2.4 Hash-Based Cryptography*

Hash-based cryptography utilizes the security properties of cryptographic hash functions to construct secure digital signatures and other cryptographic primitives. The security of hash-based schemes is rooted in the collision resistance and preimage resistance of hash functions, making them inherently resistant to quantum attacks [14]. A notable example of a hash-based cryptographic scheme is SPHINCS+, which is a stateless hash-based signature scheme that offers long-term security guarantees with minimal reliance on external assumptions [8]. It provides strong security with efficient signing and verification processes. SPHINCS+ was selected for standardization after the third-round evaluation in the NIST PQC Standardization Process and is now formally recognized as 'SLH-DSA' after standardization in 2024 [55].

*3.2.5 Isogeny-Based Cryptography*

Isogeny-based cryptography leverages the hardness of problems related to finding isogenies between elliptic curves, specifically in the context of supersingular elliptic curve isogeny graphs [25]. The primary hard problem in this domain is the Supersingular Isogeny Diffie-Hellman (SIDH) problem, which involves finding a secret isogeny between two given supersingular elliptic curves [2]. Isogeny-based schemes are attractive because they offer very small key sizes compared to other PQC families. The most well-known isogeny-based scheme is SIKE. SIKE was initially selected to advance to the fourth round of the NIST PQC Standardization Process. However, due to an efficient key-recovery attack that exploited vulnerabilities in the SIDH protocol [15], the SIKE development team included a detailed postscript in their fourth round submission to the NIST PQC Standardization Process, outlining the attack and its implications, to ensure the information is documented and open for future research.

### 3.3  NIST Security Requirements and Evaluation

NIST requires PQC algorithms to resist both quantum and classical attacks [56]. For encryption and key establishment, NIST aims to standardize schemes that offer "semantically secure" encryption against adaptive chosen ciphertext attacks (IND-CCA2 security), ensuring security even if an attacker can choose ciphertexts and get their plaintexts decrypted [37]. For ephemeral key exchanges, NIST considers IND-CPA security against chosen plaintext attacks [37], which is sufficient for scenarios with controlled environments and single-use keys. For digital signatures, NIST seeks schemes that are existentially unforgeable against adaptive chosen message attacks (EUF-CMA security), ensuring that an attacker cannot forge signatures even if they can obtain signatures for chosen messages [37].

Given the uncertainties in estimating post-quantum security, NIST defines broad security strength categories to facilitate performance comparisons and decision-making about key lengths [56]. These uncertainties result from the evolving nature of quantum computing, the potential for unforeseen quantum algorithms, and the challenges in predicting the future trajectory of quantum and classical computing. The security strength categories are presented in **Table 2**.

Security strength categories Level 1, 3, and 5, based on block cipher security, are particularly resistant to quantum attacks. Categories 2 and 4, defined by hash function security, are also resistant but require further research. Each category is defined by the computational resources required to break the cryptosystem, benchmarked against well-established symmetric schemes such as AES and SHA-3. Computational resources are measured using various metrics, including the number of classical elementary operations and quantum circuit size. To account for the practical limitations of long serial quantum computations, NIST introduces the parameter MAXDEPTH, which represents the maximum circuit depth for quantum attacks. Plausible values for MAXDEPTH range from $2^{40}$ logical gates, which is the approximate number of gates current quantum computing architectures are expected to perform serially in a year. This extends to $2^{64}$ logical gates, representing the number of gates classical computing architectures can perform serially in a decade. Finally, the upper bound of MAXDEPTH is $2^{96}$ logical gates, which is the approximate number of gates atomic scale qubits with speed-of-light propagation times could perform in a millennium [56].

NIST also considers additional security properties such as perfect forward secrecy, resistance to side-channel attacks, multi-key attacks, and misuse. These properties ensure that past session keys remain secure, the scheme is secure against physical attacks, multiple keys do not provide an advantage to attackers, and the scheme is robust against implementation errors [56].

Table 2: NIST Security Strength Categories and Computational Requirements [56]

| Security Strength Category | Description | Computational Resources Required |
| --- | --- | --- |
| Level 1 | Equivalent to breaking AES-128 | $2^{170}$/MAXDEPTH quantum gates or $2^{143}$ classical gates |
| Level 2 | Equivalent to breaking SHA3-256 | $2^{146}$ classical gates |
| Level 3 | Equivalent to breaking AES-192 | $2^{233}$/MAXDEPTH quantum gates or $2^{207}$ classical gates |
| Level 4 | Equivalent to breaking SHA3-384 | $2^{210}$ classical gates |
| Level 5 | Equivalent to breaking AES-256 | $2^{298}$/MAXDEPTH quantum gates or $2^{272}$ classical gates |

### 3.4 NIST Fourth-Round Finalists: A Comparative Overview

**Table 3** summarizes the claimed security levels, key sizes, and other relevant metrics for the NIST fourth-round finalists in the PQC Standardization Process, including the three schemes that are being standardized: CRYSTALS-Kyber, CRYSTALS-Dilithium, and SPHINCS+. All public-key encryption schemes and KEMs (i.e., CRYSTALS-Kyber, Classic McEliece, BIKE, and HQC) in **Table 3** provide IND-CCA2 security [2]. Similarly, all digital signature schemes (i.e., CRYSTALS-Dilithium, Falcon, and SPHINCS+) provide EUF-CMA security [2].

Table 3: Claimed Security Levels and Key Sizes of NIST Fourth-Round Finalists (Adapted from [2])

| PQC Candidate | Claimed NIST Security Category | Public Key (in bytes) | Private Key (in bytes) | Ciphertext/ Signature (in bytes) | Core SVP Estimate | Gate Count | Memory |
| --- | --- | --- | --- | --- | --- | --- | --- |
| CRYSTALS-Kyber512[a] | Level 1 | 800 | 1632 | 768 | C:118 bits Q:107 bits | $2^{151}$ | $2^{94}$ |
| CRYSTALS-Kyber768[a] | Level 3 | 1184 | 2400 | 1088 | C:183 bits Q:166 bits | $2^{215}$ | $2^{139}$ |
| CRYSTALS-Kyber1024[a] | Level 5 | 1568 | 3168 | 1568 | C:256 bits Q:232 bits | $2^{287}$ | $2^{190}$ |
| Classic McEliece348864 | Level 1 | 261120 | 6492 | 128 | NA | NA | NA |
| Classic McEliece460896 | Level 3 | 524160 | 13608 | 188 | NA | NA | NA |
| Classic McEliece6688128 | Level 5 | 104992 | 13932 | 240 | NA | NA | NA |
| Classic McEliece6960119 | Level 5 | 1047319 | 13948 | 226 | NA | NA | NA |
| Classic McEliece8192128 | Level 5 | 1357824 | 14120 | 240 | NA | NA | NA |
| BIKE | Level 1 | 1540 | 280 | 1572 | NA | NA | NA |
|  | Level 3 | 3082 | 418 | 3114 | NA | NA | NA |
|  | Level 5 | 5122 | 580 | 5154 | NA | NA | NA |
| HQC-128 | Level 1 | 2249 | 40 | 4481 | NA | NA | NA |
| HQC-192 | Level 3 | 4522 | 40 | 9026 | NA | NA | NA |
| HQC-256 | Level 5 | 7245 | 40 | 14469 | NA | NA | NA |
| CRYSTALS-Dilithium[a] | Level 2 | 1312 | 2528 | 2420 | C:123 bits Q:112 bits | $2^{159}$ | $2^{98}$ |

| PQC Candidate | Claimed NIST Security Category | Public Key (in bytes) | Private Key (in bytes) | Ciphertext/ Signature (in bytes) | Core SVP Estimate | Gate Count | Memory |
|---|---|---|---|---|---|---|---|
| | Level 3 | 1952 | 4000 | 3293 | C:182 bits Q:165 bits | $2^{217}$ | $2^{139}$ |
| | Level 5 | 2592 | 4864 | 4595 | C:252 bits Q:229 bits | $2^{285}$ | $2^{187}$ |
| Falcon-512[b] | Level 1 | 897 | 7553 | 666 | C:120 bits Q:108 bits | NA | NA |
| Falcon-1024[b] | Level 5 | 1793 | 13953 | 1280 | C:273 bits Q:248 bits | NA | NA |
| SPHINCS+-128s[a] | Level 1 | 32 | 64 | 7856 | NA | NA | NA |
| SPHINCS+-128f[a] | Level 1 | 32 | 64 | 17088 | NA | NA | NA |
| SPHINCS+-192s[a] | Level 3 | 48 | 96 | 16224 | NA | NA | NA |
| SPHINCS+-192f[a] | Level 3 | 48 | 96 | 35664 | NA | NA | NA |
| SPHINCS+-256s[a] | Level 5 | 64 | 128 | 29792 | NA | NA | NA |
| SPHINCS+-256f[a] | Level 5 | 64 | 128 | 49856 | NA | NA | NA |

**Note:** [a]Standardized by NIST on August 13, 2024; [b]Standardization process ongoing; C: Classical; Q: Quantum; SVP: Short Vector Problem; NA: Not Available

These public-key encryption/KEMs finalists in **Table 3** demonstrate a range of security levels from Level 1 to Level 5, with corresponding increases in key sizes and computational requirements. The key sizes and ciphertext lengths vary significantly among the schemes, reflecting different underlying cryptographic approaches and security trade-offs. For instance, lattice-based schemes, such as CRYSTALS-Kyber, have relatively small key sizes and ciphertexts, making them feasible for practical implementation, especially in resource-constrained environments. CRYSTALS-Kyber provides security levels ranging from Level 1 to Level 5, with its largest variant, CRYSTALS-Kyber1024, offering the highest security level but requiring larger keys and ciphertexts. The Core SVP estimates for CRYSTALS-Kyber range from 118 bits classically (107 bits quantumly) for CRYSTALS-Kyber512, to 256 bits classically (232 bits quantumly) for CRYSTALS-Kyber1024, indicating robust security across all levels. It is to be noted that the Core SVP Estimate measures the difficulty of finding the shortest non-zero vector in a lattice, which is a central problem in lattice-based cryptography [58]. The higher the Core SVP estimate, the more secure the scheme is. Quantum gate counts increase with the security level, from $2^{151}$ for CRYSTALS-Kyber512 to $2^{287}$ for CRYSTALS-Kyber1024, reflecting their resistance to quantum attacks. Similarly, memory requirements increase from $2^{94}$ units for CRYSTALS-Kyber512 to $2^{190}$ units for CRYSTALS-Kyber1024, indicating the trade-offs between security and resource usage. Code-based schemes, such as Classic McEliece, provide robust security with significantly larger key sizes compared to lattice-based schemes. Despite its large public key sizes, Classic McEliece's small ciphertexts make it suitable for scenarios where bandwidth is a concern. BIKE and HQC, both code-based schemes, offer a balance between security and efficiency. BIKE's security ranges from Level 1 to Level 5, with moderate key sizes and ciphertexts. HQC also spans from Level 1 to Level 5, featuring compact private keys and moderate public keys and ciphertext sizes.

Among digital signature schemes, CRYSTALS-Dilithium, which is based on lattice problems, provides security levels up to Level 5. For example, at Level 5, CRYSTALS-Dilithium has a Core SVP estimate of 252 bits classically and 229 bits quantumly, with a gate count of $2^{285}$ and a memory requirement of $2^{187}$. Falcon, another lattice-based scheme, emphasizes compact signatures and strong security guarantees, making it suitable for applications where signature size is critical. Falcon-1024, at Level 5, offers a core SVP estimate of 273 bits classically and 248 bits quantumly. SPHINCS+ provides security levels up to Level 5, with both small and large signature variants to accommodate different application needs.

The selection of a PQC scheme requires a careful evaluation of specific application requirements, including security level, efficiency, and resource constraints. The data presented in **Table 3** aids in making informed decisions by comparing the trade-offs between different schemes. The ongoing NIST PQC standardization process is important for developing robust cryptographic solutions to safeguard against the quantum computing threat. By understanding the diverse range of PQC schemes and their characteristics, organizations can make informed decisions to protect sensitive data in the quantum era.

## 4 CASE STUDY ON CRYPTOGRAPHIC VULNERABILITIES IN A TCPS APPLICATION

This section aims to illustrate the vulnerabilities of current cryptographic systems in TCPS with a case study. By examining one of the Traffic Management (TM) service packages, i.e., TM10: Electronic Toll Collection (ETC), from the ARC-IT version 9.2 [74], we highlight the potential risks posed by quantum computing to existing cryptographic mechanisms and underscore the necessity of transitioning to PQC. This analysis demonstrates how quantum attacks could exploit weaknesses in traditional cryptographic algorithms using threat modeling in the Microsoft Threat Modeling Tool 2017 [48] and provides insights into mitigating these threats using PQC solutions.

**Figure 1** illustrates the data flow and communication architecture within the TM10 service package. In our threat modeling, we focus only on the communications marked inside the black dotted box in **Figure 1**. This is because, according to the ARC-IT, the other communications depicted in **Figure 1** are initiated only if discrepancies occur during toll collection, meaning they are not activated under normal operation conditions [74]. The excluded portion involves various components, including the Traffic Management Center, Department of Motor Vehicles (DMV), Enforcement Center, Financial Center, and Fleet and Freight Management Center, which interact with the Payment Administration Center for tasks such as license requests, payment violations, financial settlements, toll data requests, and toll coordination. These components handle auxiliary functions that are critical but do not participate directly in the routine toll collection process.

Within the included portion, critical communications occur to facilitate the primary functions of electronic toll collection. The components involved are the Payment Administration Center, ITS Roadway Payment Equipment, Light Vehicle On-Board Equipment (OBE), Light Vehicle Payment Service, and Payment Device. The Payment Administration Center handles payment administration requests and disseminates payment information. It communicates with the ITS Roadway Payment Equipment to process payment instructions, authorization requests, and transactions. The ITS Roadway Payment Equipment serves as an intermediary, ensuring secure transfer of payment information and authorizing payments based on the data received. The Light Vehicle OBE is responsible for updating vehicle payment information and securing payment transactions, which it performs by interacting closely with the Light Vehicle Payment Service. This service manages toll and parking payments. The Payment Device, which could be a smart card, smartphone, or other electronic payment support devices, interfaces with the Light Vehicle OBE to actuate secure payments and update payment statuses.

The red arrows in **Figure 1** represent communications/data flow that require encryption and authentication to ensure secure and reliable transactions.

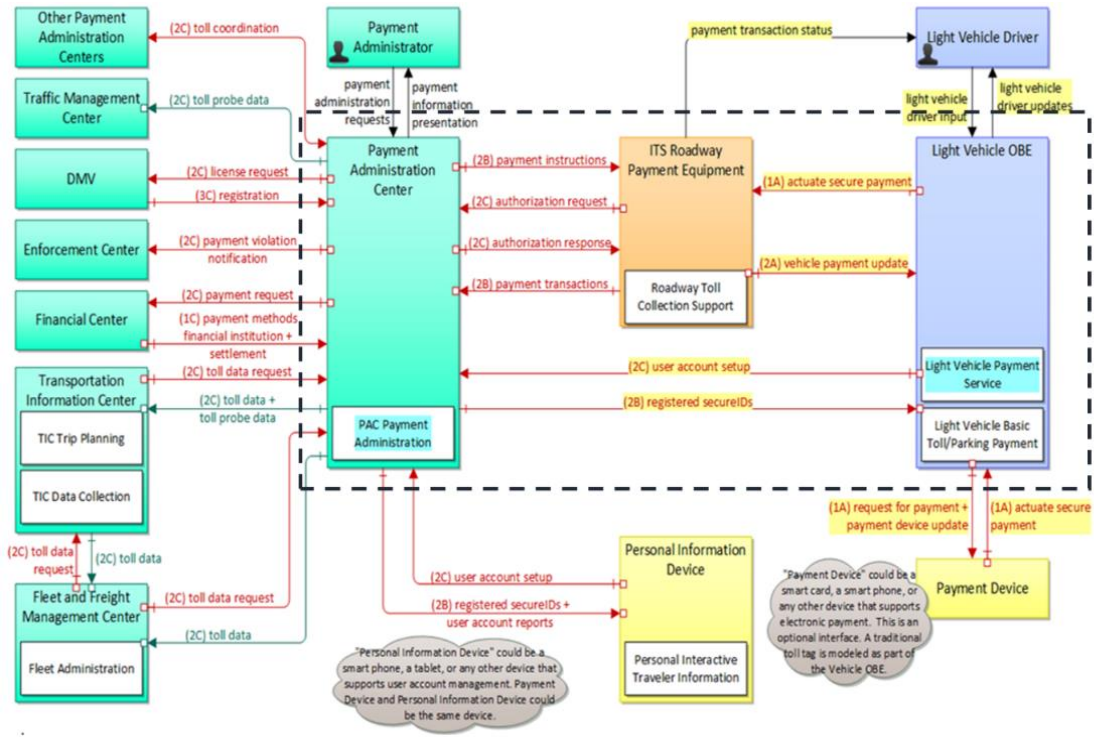

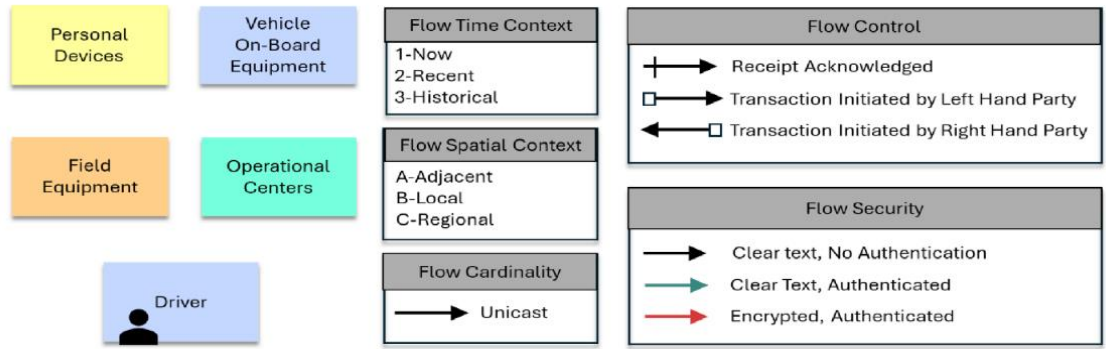

Figure 1: Architecture and data flow of TM10: Electronic Toll Collection [74]

We implemented the communications marked inside the black dotted box of **Figure 1** using the Microsoft Threat Modeling Tool [48]. **Figure 2** illustrates the data flow between the Payment Administrative Center, ITS Roadway Equipment, and Light Vehicle OBE of the ETC system using the Microsoft Threat Modeling Tool. The tool generated a report on potential threats, summarized in **Table 4**. In **Figure 2**, the 'Trust Border Boundary' delineates the perimeter within which all components and communications are considered secure and trusted, assuming that internal threats are minimal or controlled. In contrast, the 'Trust Line Boundary' indicates the transition between trusted and untrusted zones, where external threats might be more likely to penetrate or where data is more susceptible to interception and tampering. The Payment Administrative Center and ITS Roadway Payment Equipment are situated within the Trust Border Boundary, implying that communications between these components are generally secure. However, the Light Vehicle OBE, which lies outside the Trust Border Boundary and is connected through wireless communications, crosses the Trust Line Boundary. This indicates a higher risk area where data transmitted between the Light Vehicle OBE and the other components may be vulnerable to external attacks.

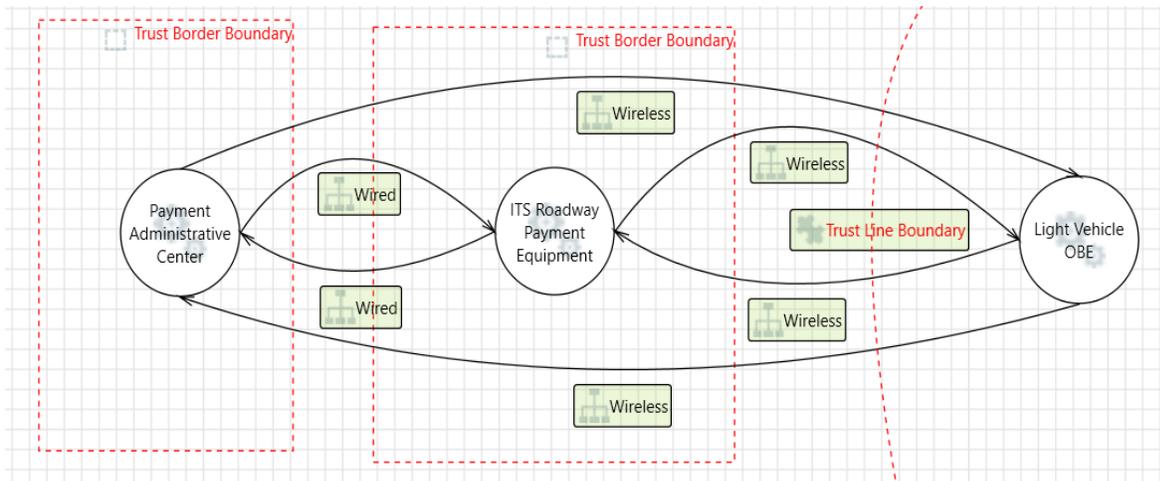

Figure 2: Data Flow between the Payment Administrative Center, ITS Roadway Equipment, and Light Vehicle OBE using Microsoft Threat Modeling Tool [48]

Table 4: Summary of Microsoft Threat Modeling Tool Report

| Title | Category | Interaction | Priority |
| --- | --- | --- | --- |
| Weak Authentication Scheme | Information Disclosure | Wired | High |
| Collision Attacks | Tampering | Wired | High |
| Replay Attacks | Tampering | Wired | High |
| Elevation Using Impersonation | Elevation of Privilege | Wired | High |
| ITS Roadway Payment Equipment May be Subject to Elevation of Privilege Using Remote Code Execution | Elevation of Privilege | Wireless | High |
| Elevation by Changing the Execution Flow in Light Vehicle OBE | Elevation of Privilege | Wireless | High |
| ITS Roadway Payment Equipment May be Subject to Elevation of Privilege Using Remote Code Execution | Elevation of Privilege | Wired | High |
| Elevation by Changing the Execution Flow in ITS Roadway Payment Equipment | Elevation of Privilege | Wired | High |
| Payment Administrative Center May be Subject to Elevation of Privilege Using Remote Code Execution | Elevation of Privilege | Wireless | High |

| Title | Category | Interaction | Priority |
|---|---|---|---|
| Elevation by Changing the Execution Flow in Payment Administrative Center | Elevation of Privilege | Wireless | High |
| Light Vehicle OBE May be Subject to Elevation of Privilege Using Remote Code Execution | Elevation of Privilege | Wireless | High |

**Table 4** highlights vulnerabilities that are all highly susceptible to quantum attacks. These vulnerabilities are critical because quantum computing can efficiently break the underlying mechanism of cryptographic techniques that are currently available. For example, weak authentication schemes and collision attacks have become feasible with the advent of quantum computing capabilities [12, 14]. **Table 4** also highlights various threat categories, such as information disclosure, tampering, elevation of privilege, denial of service, and repudiation, indicating a high priority for mitigation. Elevation of privilege using impersonation, remote code execution, and changing the execution flow in ITS Roadway Payment Equipment and Light Vehicle OBE are among the most severe threats identified.

**Table 5** elaborates on the vulnerabilities identified in **Table 4**, specifically highlighting those susceptible to quantum attacks. For each vulnerability, **Table 5** describes the potential quantum attack that could be exploited, the impact of such an attack, and the suggested mitigation using PQC techniques. For example, weak authentication schemes are vulnerable to quantum attacks using Shor's algorithm, necessitating the adoption of lattice-based PQC algorithms, such as CRYSTALS-Kyber [13] or Falcon [80]. Collision attacks can be mitigated by hash-based PQC algorithms, such as SPHINCS+ [8], to prevent unauthorized data modifications. Replay attacks, enhanced by quantum computing, can be countered with time-based PQC protocols to ensure transaction uniqueness. Elevation using impersonation requires strong identification verification mechanisms provided by lattice-based PQC schemes to prevent unauthorized access. Finally, elevation by changing the execution flow can be mitigated through robust quantum-resistant algorithms and secure coding practices to maintain system integrity.

Table 5: Vulnerabilities, Quantum Attacks, Impacts, and Suggested Mitigations for Electronic Toll Collection System

| Vulnerability | Potential Quantum Attack | Impact | Suggested Mitigation |
|---|---|---|---|
| Weak Authentication Scheme | Shor's algorithm can efficiently factorize large integers and solve discrete logarithms, breaking RSA and ECC. | Unauthorized access and breached confidentiality | Use PQC algorithms, such as CRYSTALS-Kyber or Falcon |
| Collision Attacks | Grover's algorithm can find the pre-image of a hash function with a quadratic speedup, reducing the security level. | Data tampering | Use hash-based PQC algorithms like SPHINCS+ |
| Replay Attacks | With compromised cryptographic protocols, attackers can intercept and retransmit valid data packets. | System manipulation, repeated transactions, and fraud | Implement time-based protocols and secure communication channels with PQC |
| Elevation Using Impersonation | Attackers can impersonate legitimate users or devices by breaking the underlying cryptographic keys. | Unauthorized actions and security breaches | Adopt lattice-based PQC schemes (i.e., CRYSTALS-Kyber) |
| Elevation by Changing Execution Flow | If the encryption used to secure communication is compromised, attackers can insert malicious code. | System takeover and data breaches | Secure communication with PQC algorithms and implement robust input validation and security checks (secure PQC coding practices) |

In conclusion, this case study underscores the pressing need to transition to PQC to secure critical transportation infrastructures against the impending threat of quantum computing. By adopting PQC solutions, we can safeguard the integrity, confidentiality, and authenticity of essential TCPS applications, such as electronic toll collection, ensuring their resilience in a post-quantum world.

## 5 EXPERIMENTAL EVALUATION AND SECURITY ASSESSMENT OF CRYSTALS-KYBER FOR TCPS

This section presents a comprehensive evaluation of CRYSTALS-Kyber (Kyber), focusing on its applicability and effectiveness as a post-quantum cryptographic mechanism for TCPS. Kyber has been selected for this evaluation for its robust security framework, which meets the strict demands of TCPS to ensure the integrity, confidentiality, and authenticity of critical data exchanges. Standardized by NIST on August 13, 2024, as the "Module-Lattice-Based Key-Encapsulation Mechanism," Kyber is designed to secure communication against both classical and quantum attacks [5]. Kyber's foundational reliance on the Module Learning with Errors (MLWE) problem ensures its resistance to quantum algorithms such as Shor's Algorithm, which compromise traditional cryptographic schemes, such as RSA and ECC. Additionally, its compliance with advanced security standards, such as IND-CCA2, highlights its ability to prevent adaptive chosen ciphertext attacks, which is essential for maintaining the integrity of TCPS communication networks.

To analyze Kyber's effectiveness comprehensively, this section integrates both the security and performance perspectives. The following subsections begin by discussing Kyber's critical security features to support TCPS, including its foundational strengths, resistance to known attack vectors, and a discussion on the quantum resource estimates required to breach its defenses. Subsequently, the experimental setup and simulation scenarios used to evaluate Kyber's performance across diverse peer-to-peer (P2P) communication networks in TCPS are detailed. This comprehensive evaluation demonstrates Kyber's feasibility as a foundation of PQC for securing critical TCPS applications.

### 5.1 Security Features of Kyber to Support TCPS

*5.1.1 Fundamental Security Features*

Kyber's security derives from the hardness of the MLWE problem, a lattice-based problem resistant to classical and quantum attacks [13]. Unlike RSA and ECC, which are vulnerable to Shor's algorithm, Kyber's lattice-based structure remains secure due to its inherent complexity of solving lattice-based problems. This complexity is further reinforced by Kyber's carefully chosen parameter sets, which balance cryptographic strength, computational efficiency, and decryption failure probability. These parameter sets play a crucial role in ensuring Kyber's robustness against cryptanalytic attacks, including brute-force attack accelerated by quantum algorithms, such as Grover's algorithm [5].

Kyber defines three parameter sets for three different Kyber variants, i.e., Kyber-512, Kyber-768, and Kyber-1024, corresponding to NIST security levels 1, 3, and 5, respectively (**Table 6**). Each parameter set specifies the following critical values [5]:

- $n$: Polynomial degree, fixed at 256 for all variants, which defines the size of the polynomial ring used in Kyber. This choice ensures computational efficiency and high levels of security.
- $k$: Number of polynomials in the key and ciphertext. This parameter scales with the desired security level, with higher values of $k$ providing greater cryptographic strength (e.g., $k = 2$ for Kyber-512 and $k = 4$ for Kyber-1024).
- $q$: Modulus for polynomial coefficients, fixed at 3329 for all variants. This modulus prevents coefficient overflow while preserving the hardness of the MLWE problem.

- $\eta$: Error distribution parameter that introduces noise into the MLWE framework, adding cryptographic strength by obscuring the relationship between plaintext and ciphertext.
- $\tau$: Number of noise terms sampled during the key encapsulation process, ensuring additional randomness and robustness against attacks.
- $d_u, d_v$: Bit lengths for compressing the ciphertext components $u$ and $v$. These compression parameters optimize the storage and transmission of ciphertexts without compromising security.
- $\delta$: Decryption failure probability, representing the likelihood of decryption errors. Kyber's parameter sets ensure that $\delta$ remains negligible, ensuring compliance with stringent security standards.

Table 6: Parameter Sets for Kyber [5]

| Kyber variant | $n$ | $k$ | $q$ | $\eta$ | $\tau$ | $(d_u, d_v)$ | $\delta$ |
|---|---|---|---|---|---|---|---|
| Kyber-512 | 256 | 2 | 3329 | 3 | 2 | (10, 4) | $2^{-139}$ |
| Kyber-768 | 256 | 3 | 3329 | 2 | 2 | (10, 4) | $2^{-164}$ |
| Kyber-1024 | 256 | 4 | 3329 | 2 | 2 | (11, 5) | $2^{-174}$ |

Kyber's use of the above parameter sets ensures its compliance with high standards for IND-CPA and IND-CCA2 security [5]. CPA security ensures that attackers cannot infer plaintext information from the ciphertext, protecting initial key exchanges. CCA security protects against active attacks where manipulated ciphertexts could otherwise expose the system to vulnerabilities. These properties are critical for TCPS, ensuring data integrity and authenticity in dynamic environments susceptible to tampering and eavesdropping.

*5.1.2 Core SVP Estimates, Resistance to Primal and Dual Attacks, and Quantum Resource Requirements for Kyber*

The security of Kyber is further quantified through Core SVP estimates, which indicate the computational difficulty of solving its lattice problems. These estimates are derived based on the complexity of primal and dual attacks, which are the two main methods used to attack lattice-based cryptographic schemes [4, 40, 79]. Primal attacks focus on directly recovering the secret key by finding the shortest vector in the lattice that corresponds to the error term in Kyber's MLWE framework. Advanced lattice reduction techniques, like the Block Korkine-Zolotarev (BKZ) algorithm and lattice sieving, are employed to approximate this vector [22, 43, 49, 72]. Dual attacks, on the other hand, aim to distinguish between encrypted vectors and random noise by exploiting structural discrepancies in the lattice. Techniques, such as Fourier analysis and lattice sieving, are employed to construct distinguishers capable of isolating distinctive distributions within the MLWE framework [3, 18].

The Core SVP estimates for Kyber-512, Kyber-768, and Kyber-1024 (**Table 3**) reflect their resistance to known attacks, quantifying the computational effort required to compromise their security. For instance, Kyber-512 has a Core SVP estimate of 118 bits in the classical setting and 107 bits in the quantum setting, requiring approximately $2^{118}$ classical or $2^{107}$ quantum operations to compromise its security. Higher variants, such as Kyber-1024 offer even greater security margins, i.e., Core-SVP estimates of 256 bits in a classical setting and 236 bits in a quantum setting, making them ideal for TCPS where stronger security is critical.

Furthermore, these Core SVP estimates align closely with NIST-defined security levels for PQC schemes (**Table 2**), which are benchmarked against the equivalent security provided by AES. Specifically, NIST Security Level 1 corresponds to the security of AES-128, Level 3 to AES-192, and Level 5 to AES-256. Kyber-512, Kyber-768, and Kyber-1024 are designed to meet these security levels, respectively, offering protection analogous to their AES counterparts. For example, based on studies by Gheorghiu and Mosca (2025), a quantum attack using Grover's algorithm on AES-256, which provides

NIST Security Level 5, would require approximately $2^{240}$ physical qubits and $2^{205}$ quantum CPUs operating in parallel for one year, assuming a very low physical error rate [27]. Similarly, AES-192 (NIST Security Level 3) demands approximately $2^{165}$ physical qubits and $2^{140}$ quantum CPUs, while AES-128 (Security Level 1) requires approximately $2^{100}$ physical qubits and $2^{80}$ quantum CPUs for the same attack duration [27]. Mapping these requirements to Kyber, the resource demands for compromising Kyber-1024 align with AES-256, demonstrating Kyber's robustness at Security Level 5 against quantum adversaries. Similarly, Kyber-768 and Kyber-512 reflect the computational challenges associated with AES-192 and AES-128, respectively, aligning with their respective security levels.

However, current quantum computing capabilities fall significantly short of these requirements. For instance, IBM's quantum roadmap and similar efforts by Google and Quantinuum project quantum systems scaling to thousands of qubits over the coming years [88–90]. Despite these advancements, the projected systems remain orders of magnitude below the $2^{240}$ physical qubits and $2^{205}$ quantum CPUs needed to compromise Kyber-1024. Even for Kyber-512, the minimum physical and computational requirements far exceed the capabilities of existing or near-term quantum computers. This contrast between resource requirements and current projections highlights the immense resilience of Kyber against foreseeable quantum threats. As a result, Kyber remains a robust and practical PQC solution for securing critical systems like TCPS, where long-term data protection is paramount.

*5.1.3 Resistance to Algebraic and Symmetric Primitive Attacks*

Kyber also exhibits strong resilience to algebraic attacks [5]. Quantum algorithms targeting the Ideal Shortest Vector Problem (Ideal-SVP) [11, 19] face significant computational challenges when applied to Kyber's MLWE structure, which is more complex than simpler lattice types like Ring-LWE. The additional complexity ensures that Kyber's cryptographic foundation remains secure against these advanced algebraic attacks [5]. Kyber's use of symmetric functions, particularly Keccak-derived algorithms, such as SHAKE-128 and SHAKE-256 [10], enhances its resistance to cryptanalysis by providing strong pseudo-randomness, high resistance to differential and linear attacks, and robustness against quantum adversaries due to their well-studied and secure design [5]. These functions are critical for key encapsulation and pseudorandom generation, providing robust protection against both classical and quantum adversaries.

*5.1.4 Protection against Decryption Failures*

Kyber minimizes decryption failure probabilities, a key aspect of maintaining CCA security [5]. Low decryption failure rates reduce the risk of attackers exploiting modified ciphertexts to infer secret keys. For example, Kyber-512 has a decryption failure probability of approximately $2^{-69}$ ($2^{-82}$ for Kyber-768 and $2^{-87}$ for Kyber-1024) under quantum assumptions [5, 13]. Specifically, this means if an attacker attempts to exploit this probability, they would need to submit $2^{69}$ ciphertexts to statistically observe at least one decryption failure that might reveal sensitive information, ensuring that Grover's algorithm cannot feasibly exploit this vulnerability. These design choices reinforce Kyber's reliability in high-stakes environments like TCPS, where secure communication is essential.

For TCPS, this assessment underscores the strong security features and robustness of Kyber's lattice-based cryptographic mechanisms. Furthermore, the immense physical qubit and quantum processor requirements for breaking Kyber, which remain far beyond the current capabilities of quantum technology, suggest that Kyber could remain a secure PQC option for encrypting sensitive data and ensuring authentication within TCPS applications. This includes safeguarding communication in ITS and V2X networks, where low-latency and high-reliability security protocols are paramount. Thus, the alignment of Kyber's inherent security features, resistance to known attack vectors, and compatibility with the projected limitations of quantum computing reinforces its viability as a quantum-resistant solution for critical infrastructure protection in TCPS. To further evaluate Kyber's practical applicability, the next section delves into the simulation scenarios and experimental setup used to test its performance in diverse P2P communication networks within TCPS. These simulations provide critical insights into Kyber's operational feasibility in real-world environments.

### 5.2 Simulation Scenarios and Setup

For the simulation of the Kyber KEM within a P2P communication network, Objective Modular Network Testbed in C++, also known as OMNeT++ [91] was selected as the simulation platform. OMNeT++ offers a modular and extensible architecture that is well-suited for implementing custom cryptographic protocols. Furthermore, the platform's scalability and robust visualization tools enable detailed performance analysis, which is essential for evaluating the feasibility of Kyber in practical communication network scenarios. We implemented the INET framework [92] to our OMNeT++ environment to provide a comprehensive suite of internet protocols, which enabled realistic modeling of both wired and wireless P2P networks. Additionally, SimuLTE [76] was implemented to simulate LTE-based wireless communication scenarios, allowing for the evaluation of Kyber in wireless cellular environments. Specifications of simulation parameters are presented in **Table 7**.

Table 7: Simulation Environment Setup

| Parameter | Details/Specification |
| --- | --- |
| Main Machine | Intel(R) Core (TM) i7-7700 CPU @ 3.60GHz |
| Virtual Machine | Instant-veins-5.2-i1 |
| Operating System (OS) of the Virtual Machine | Debian 11, Linux 5, GNOME 3 |
| Number of CPU for the Main Machine | 8 |
| Number of CPU assigned to the Virtual Machine | 8 |
| RAM size for Main Machine | 32 Gigabytes (GB) |
| RAM size for Virtual Machine | 24 GB |
| Time per clock cycle | 0.29 nanoseconds (ns) |
| OMNeT++ version | 5.7 |
| INET Framework version | 4.2.8 |
| SimuLTE version | 1.2.0 |

We simulated four different scenarios to analyze the performance and applicability of Kyber in TCPS P2P communication. First, we evaluated wired communication between two static nodes. This represents the wired Ethernet communication between two stationary nodes in TCPS, for example, the communication between the Payment Administration Center and DMV in ARC-IT service package TM10: Electronic Toll Collection, as shown in **Figure 1**. Then, we evaluated wireless communication between two static nodes. We can consider the communication between ITS Roadway Payment Equipment and the Payment Administration Center in TM10 as an example of this. After that, we evaluated wireless communication between one static and one dynamic node. We can consider the communication between ITS Roadway Payment Equipment and a moving vehicle as an example of this. Finally, we evaluated wireless communication between two dynamic nodes, which can be depicted as the communication between two moving vehicles, for example exchanging basic safety messages (BSMs) between two connected vehicles. These four scenarios depict all the possible P2P communication in TCPS and are illustrated in **Figure 3**.

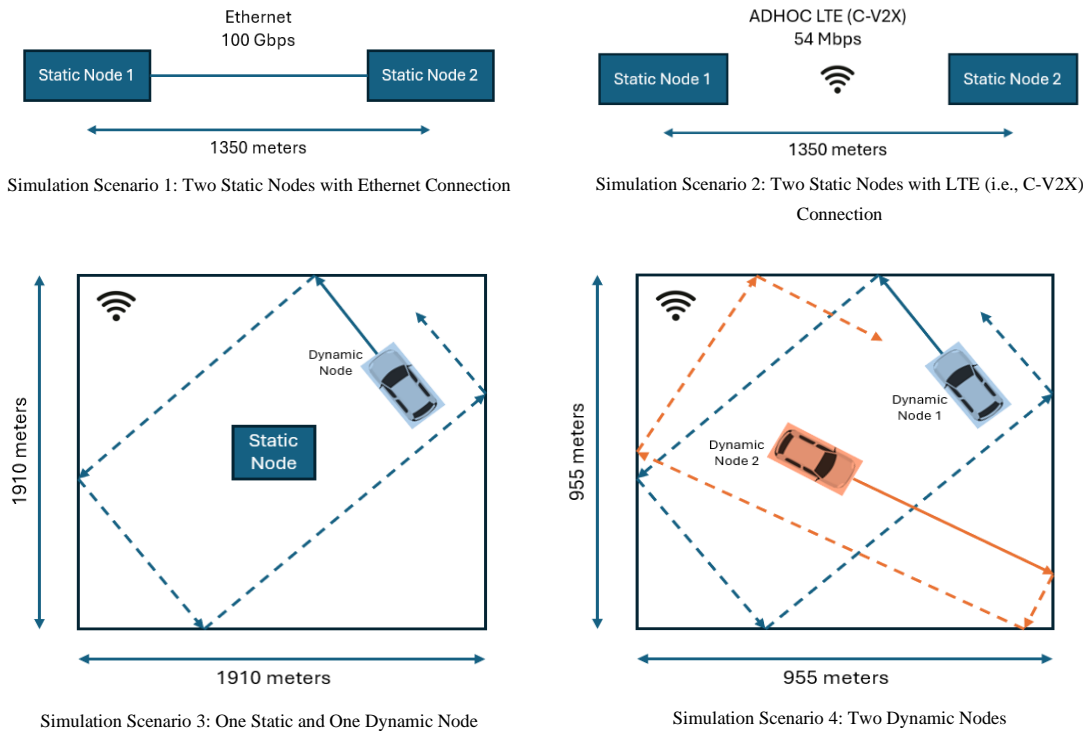

Figure 3: Simulation Scenarios for Evaluating Kyber in Wired and Wireless Communication Networks in TCPS

The distances and mobility configurations for each scenario were designed to align with real-world TCPS communication requirements. For scenarios 1 and 2, involving two static nodes, the distance between the nodes was set to 1350 meters, reflecting the typical communication range of Cellular Vehicle-to-Everything (C-V2X) technology [60]. While this distance is not a limiting factor for Ethernet-based communication due to its high bandwidth and low-latency characteristics, we adopted the same configuration to maintain consistency across all scenarios and facilitate a fair comparison between wired and wireless setups. In scenarios 3 and 4, where dynamic nodes were introduced, a linear waypoint mobility model was employed to simulate their movement within defined bounding boxes [93]. Each dynamic node moved at a constant speed of 40 meters per second, a value representative of typical freeway speeds.

For simulation scenario 3, the static node was placed at the center of a bounding box with sides measuring 1910 meters. This configuration ensured that the maximum distance between the static and dynamic nodes was 1350 meters, occurring when the dynamic node was positioned at the corner of the bounding box. In simulation scenario 4, the bounding box was reduced to 955 meters on each side, ensuring that the maximum distance between the two dynamic nodes remained 1350 meters when they occupied opposite corners of the box. These configurations allowed for a realistic evaluation of communication performance under varying mobility and distance constraints.

For wired communication, we utilized the Ethernet with a bandwidth of 100 Gigabits per second (Gbps), providing a high-speed, low-latency network suitable for modern TCPS infrastructure. In our wireless communication simulations, we employed ad hoc LTE with a bandwidth of 54 Megabits per second (Mbps). This choice aligns with data rates observed in real-world LTE networks under typical operating conditions. Ad hoc LTE represents peer-to-peer C-V2X communication, enabling direct communication between vehicles and other devices without relying on network infrastructure. For instance, the IEEE 802.11g standard, operating in the 2.4 GHz band, supports a maximum data rate of 54 Mbps, which is commonly utilized in ad hoc network configurations [31]. Additionally, studies have shown that LTE networks can achieve data rates up to 54 Mbps in various scenarios, including vehicular networking applications [29]. These data rates are consistent with moderate network congestion and typical signal strengths found in urban and suburban environments.

To implement the cryptographic operations, the liboqs library [94] was employed for Kyber KEM, and OpenSSL [95] was utilized for AES-256 for symmetric key encryption and decryption. First, a shared secret key was established between the two peers using Kyber. One peer generated a Kyber public-private key pair and shared the public key with the other peer, while keeping the private key securely stored. The second peer used the received public key to generate a ciphertext and a shared secret key. The ciphertext was then transmitted back to the first peer. Upon receiving the ciphertext, the first peer used its private key to recover the shared secret key, ensuring that both peers possessed the same shared secret key. After establishing the shared secret key, each peer utilized that key to encrypt and decrypt data (32-byte message) with AES-256 symmetric cryptographic scheme. AES-256 was chosen because, even if an attacker with a quantum computer attempts to decrypt the ciphertext using Grover's algorithm, AES-256 could still provide 128 bits of security (i.e., $2^{128}$ brute-force iterations to break the ciphertext).

The simulation was conducted with all three Kyber variants, e.g., Kyber-512, Kyber-768, and Kyber-1024, to analyze the computational and communication overhead associated with different key sizes. Key sizes and ciphertext payload sizes were matched to the specifications of Kyber's three variants (see **Table 3**) for consistency. To simplify the analysis, the effect of packet fragmentation was not considered. Instead, payload sizes were set to correspond directly to the public key and ciphertext sizes of each cryptographic scheme.

### 5.3 Performance Evaluation

The performance evaluation was conducted by simulating Kyber's KEM in both wired Ethernet and wireless ad hoc LTE (i.e., C-V2X) network scenarios, which are commonly encountered in TCPS. Two primary performance metrics were analyzed: (1) execution time for cryptographic operations and (2) communication delays across different network media. For execution time metrics (**Table 8**), average values from five simulation runs were reported to account for variability. For communication delay metrics (**Table 9**), maximum, minimum, and average values from the simulation runs were recorded to provide a comprehensive view of performance under varying conditions.

**Table 8** presents the execution times for key cryptographic operations, i.e., key generation, encapsulation, and decapsulation, for each variant of Kyber (Kyber-512, Kyber-768, and Kyber-1024). The data indicate that execution time increases with the complexity of the cryptographic variant. For instance, Kyber-512 requires 44 microseconds (µs) for key generation, while Kyber-1024 requires 107 µs. Similarly, encapsulation times range from 53 µs for Kyber-512 to 121 µs for Kyber-1024, and decapsulation times range from 65 µs to 147 µs across the three variants. This increase in execution time is primarily attributed to the larger key and ciphertext sizes in higher variants, which necessitate additional computational steps for cryptographic operations.

Table 8: Execution time for cryptographic operations

| PQC Scheme | Key Generation (cycle count/time ($\mu s$)) | Encapsulation (cycle count /time ($\mu s$)) | Decapsulation (cycle count /time ($\mu s$)) |
| --- | --- | --- | --- |
| Kyber-512 | 155365 / 44 | 191358 / 53 | 232691 / 65 |
| Kyber-768 | 272804 / 75 | 320600 / 89 | 365776 / 101 |
| Kyber-1024 | 387324 / 107 | 436250 / 121 | 531310 / 147 |

**Table 9** presents the communication delays incurred when transmitting public keys, ciphertexts, and encrypted data over the Ethernet and the ad hoc LTE network (i.e., C-V2X) mediums. In the Ethernet medium, communication delays are minimal and consistent across all PQC schemes. Public Key and Encrypted Data transmission delays typically average around 5 µs, while Ciphertext transmission takes up to 10 µs. These results indicate that in a wired network with high bandwidth and low latency, the variation in key sizes and cryptographic operations have a negligible impact on communication delays.

Table 9: Communication delay (in $\mu s$) for different simulation scenarios

| PQC Scheme | Simulation Scenario | Medium | | Public Key | Ciphertext | Encrypted Data |
|---|---|---|---|---|---|---|
| Kyber-512 | Static-Static | Ethernet | Maximum (max) | 5.0029 | 10.0078 | 5.0056 |
| | | | Minimum (min) | 5.0023 | 10.0056 | 5.0031 |
| | | | Average (avg) | 5.0027 | 10.0060 | 5.0045 |
| | Static-Static | ADHOC LTE (C-V2X) | max | 1,126.13 | 1,001,948 | 676.10 |
| | | | min | 1,125.97 | 1,001,948 | 676.03 |
| | | | avg | 1,126.05 | 1,001,948 | 676.05 |
| | Static-Dynamic | ADHOC LTE | max | 1,135.15 | 1,001,960 | 685.15 |
| | | | min | 1,122.40 | 1,001,941 | 672.40 |
| | | | avg | 1,129.16 | 1,001,952 | 679.16 |
| | Dynamic-Dynamic | ADHOC LTE | max | 1,374.73 | 1,002,139 | 1,564.72 |
| | | | min | 1,362.00 | 1,002,122 | 1,532.00 |
| | | | avg | 1,369.25 | 1,002,132 | 1,547.24 |
| Kyber-768 | Static-Static | Ethernet | max | 5.0035 | 10.0064 | 5.0053 |
| | | | min | 5.0021 | 10.0061 | 5.0046 |
| | | | avg | 5.0029 | 10.0063 | 5.0048 |
| | Static-Static | ADHOC LTE | max | 1,254.08 | 1,002,184 | 676.08 |
| | | | min | 1,254.02 | 1,002,183 | 676.01 |
| | | | avg | 1,254.05 | 1,002,183 | 676.05 |
| | Static-Dynamic | ADHOC LTE | max | 1,263.16 | 1,002,196 | 685.16 |
| | | | min | 1,122.40 | 1,001,943 | 672.40 |
| | | | avg | 1,206.14 | 1,002,093 | 679.33 |
| | Dynamic-Dynamic | ADHOC LTE | max | 1,502.73 | 1,002,375 | 1,672.72 |
| | | | min | 1,490.01 | 1,002,358 | 1,640.01 |
| | | | avg | 1,497.19 | 1,002,368 | 1,659.19 |
| Kyber-1024 | Static-Static | Ethernet | max | 5.0037 | 10.0064 | 5.0046 |
| | | | min | 5.0032 | 10.0056 | 5.0040 |
| | | | avg | 5.0036 | 10.0061 | 5.0045 |
| | Static-Static | ADHOC LTE | max | 1,001,477 | 1,002,661 | 676.25 |
| | | | min | 1,001,469 | 1,002,641 | 676.18 |
| | | | avg | 1,001,473 | 1,002,656 | 676.20 |
| | Static-Dynamic | ADHOC LTE | max | 1,001,492 | 1,002,685 | 685.18 |
| | | | min | 1,001,471 | 1,002,651 | 672.50 |
| | | | avg | 1,001,484 | 1,002,672 | 680.36 |
| | Dynamic-Dynamic | ADHOC LTE | max | 1,001,731 | 1,002,864 | 1,933.21 |
| | | | min | 1,001,710 | 1,002,830 | 1,892.01 |
| | | | avg | 1,001,722 | 1,002,849 | 1,915.99 |

Conversely, in the ad hoc LTE (i.e., C-V2X) network, communication delays increase significantly, particularly for larger key sizes and scenarios involving dynamic nodes. For Kyber-512 in the Static-Static scenario, the maximum delay for transmitting the Public Key is around 1,126 µs, while the Ciphertext delay peaks at approximately 1,001,948 µs. These delays escalate further in the Dynamic-Dynamic scenario, where Public Key transmission delay reaches 1,374 µs and Encrypted Data transmission delay rises to 1,564 µs.

For Kyber-768, the increased key size results in marginally higher delays. In the Static-Static LTE scenario, Public Key transmission reaches up to a maximum of 1,254 µs, with Ciphertext delays peaking at 1,002,184 µs. In the Dynamic-Dynamic scenario, Public Key delay increases to 1,502 µs, and Encrypted Data delay reaches to 1,672 µs.

The largest key size, Kyber-1024, incurs the most significant delays in ad hoc LTE networks. In the Static-Static LTE scenario, Public Key transmission peaks around 1,001,477 µs, while Ciphertext transmission peaks at 1,002,661 µs. These delays further increase in the Dynamic-Dynamic scenario, with maximum Public Key transmission delay reaching 1,001,731 µs and Encrypted Data delay peaking at 1,933 µs.

The results presented in **Table 9** demonstrate a contrast in communication performance between the wired Ethernet and the wireless ad hoc LTE network when implementing all variants of Kyber for various TCPS communication scenarios. Average communication latencies for all Kyber variants over Ethernet networks remain consistently low, with delays of approximately 5 µs ($5 \times 10^{-3}$ milliseconds [ms]) for public key and encrypted data transmissions, and around 10 µs ($10^{-2}$ ms) for ciphertext transmission. These results highlight the capability of high-bandwidth, low-latency Ethernet networks to efficiently handle the increased payload sizes associated with higher-level PQC variants, such as Kyber-1024. This makes Ethernet a suitable medium for real-time TCPS applications, including safety-critical systems requiring latencies under 100 ms, e.g., vehicle collision warning and lane change assistance, as stipulated by standards such as Society of Automative Engineers (SAE) J2735 [75] and European Telecommunications Standards Institute (ETSI) TR 102 638 [24].

In contrast, the ad hoc LTE network demonstrates significantly higher communication delays, particularly for scenarios involving larger cryptographic payloads for mobile vehicular nodes. For instance, in a static-dynamic scenario, the average public key transmission latency for Kyber-512 is approximately 1.13 ms, increasing further to 1.21 ms for Kyber-768 and exceeding 1 second for Kyber-1024. Ciphertext transmission in the same scenario experiences delays exceeding 1 second (1,000 ms) for all Kyber variants. These delays far exceed the 100-ms latency requirement for safety-critical TCPS applications, making ad hoc LTE impractical for real-time safety-critical TCPS applications such as collision warning.

Furthermore, it is important to note that the current simulation does not account for the effects of packet fragmentation, which becomes a critical factor for larger cryptographic payloads like Kyber's public keys and ciphertexts. Packet fragmentation in real-world TCPS data transmission networks would introduce additional delays due to the need for reassembly and retransmission processes, further increasing the observed communication latencies.

In summary, while high-bandwidth Ethernet networks demonstrate that it is feasible to implement Kyber to satisfy latency requirements of real-time TCPS applications, ad hoc LTE networks face significant limitations due to their inherent bandwidth constraints and higher latency. To address these challenges in wireless environments, future research should explore the integration of new-generation wireless technologies, such as 5G, which offer lower latency and greater bandwidth. Additionally, optimizing PQC schemes for reduced payload sizes, developing lightweight PQC solutions, and designing hybrid PQC schemes that integrate traditional public-key cryptography with PQC algorithms could help mitigate latency impacts, enhancing the viability of wireless PQC implementations for TCPS.

# 6 CHALLENGES AHEAD AND FUTURE DIRECTIONS

As PQC begins to be integrated into TCPS, several challenges related to its deployment and implementation have emerged. While NIST's standardization of schemes, such as CRYSTALS-Kyber, CRYSTALS-Dilithium, and SPHINCS++, marks a significant step forward, the road to widespread adoption in TCPS remains complex. The following are key challenges and potential future directions for PQC adoption in TCPS.

## 6.1 Evolving Threats

PQC algorithms, though resistant to quantum attacks, are vulnerable to physical side-channel attacks (SCA) and fault injection attacks (FIA). Lattice-based schemes, such as CRYSTALS-Kyber and CRYSTALS-Dilithium, have been shown to leak sensitive information through power consumption and electromagnetic emissions during operations like polynomial multiplication and matrix operations. For instance, correlation power analysis (CPA) attacks on CRYSTALS-Kyber have been successful in recovering secret keys by exploiting the Number Theoretic Transform (NTT) used in polynomial multiplication [32, 39, 61]. Similarly, CRYSTALS-Dilithium has been targeted by Differential Power Analysis (DPA) during its signing process, making it vulnerable to key recovery [61]. Falcon, another lattice-based signature scheme, has also been found susceptible to floating-point operation side-channel attacks, with attackers exploiting Fast Fourier Transform (FFT) operations to recover signing keys [35]. These vulnerabilities demonstrate that, despite their quantum resilience, PQC schemes are still vulnerable to physical attacks, which are becoming more sophisticated. Techniques such as side-channel analysis are increasingly enhanced by machine learning and deep learning models, improving the efficiency and success rate of these attacks [32, 62]. To address these threats, countermeasures, such as masking, hiding, and constant-time algorithms, need to be implemented across various TCPS. However, these protections often come with substantial performance overhead, especially for resource-constrained embedded systems used in vehicles [34, 61]. Future research must focus on balancing security with performance, ensuring that these schemes remain efficient while being resilient against side-channel and fault injection attacks [66].

## 6.2 Performance and Real-Time Constraints

A critical challenge in deploying PQC within TCPS is managing the increased computational load and latency introduced by PQC algorithms. Many TCPS components, particularly V2V communications, require low-latency and real-time processing to support functions, such as collision warnings and emergency braking systems [36]. Our simulation results indicate that while high-bandwidth Ethernet networks effectively handle the increased payload sizes of PQC schemes with minimal communication delays (e.g., around 20 μs [$2 \times 10^{-2}$ ms] across all Kyber variants), ad hoc LTE networks experience significant latency challenges (e.g., exceeding 1 second [1,000 ms] across all Kyber variants). Thus, in wireless environments, delays for transmitting ciphertext and encrypted data can exceed the 100-ms threshold required for safety-critical TCPS applications like collision avoidance, making the ad hoc LTE wireless network impractical for such real-time scenarios.

The larger key sizes and signature lengths of PQC schemes increase both computation time and message size, which can lead to significant performance degradation in real-time environments, particularly in wireless mediums [42, 66, 67]. To mitigate these challenges, future research should explore lightweight variants of PQC algorithms or develop new hybrid schemes that integrate traditional public-key cryptography with PQC algorithms. Such approaches could reduce payload sizes while maintaining robust security for real-time TCPS applications [67]. Additionally, optimizing hardware and software implementations to minimize latency and enhance processing efficiency, such as leveraging new-generation wireless technologies such as 5G, will be crucial in ensuring the successful deployment of PQC in performance-sensitive TCPS [34, 66].

### 6.3 Compatibility and Scalability

One of the primary challenges in adopting PQC within TCPS is ensuring compatibility with existing cryptographic infrastructure and interoperability across different platforms. Current systems rely heavily on classical cryptographic standards, such as ECC, which are deeply embedded in protocols, i.e., IEEE 1609.2 for V2X communication. Transitioning to PQC requires hybrid cryptographic solutions, where classical and post-quantum schemes are used together during the migration phase [73]. Moreover, the larger key sizes and signature lengths inherent in PQC algorithms introduce additional challenges. These larger parameters burden the latency requirements of V2X communication, such as the 100 ms threshold [24, 36, 75], which are critical for real-time TCPS applications such as collision avoidance [67, 73]. To ensure scalability, future research should focus on optimizing PQC algorithms, perhaps through compression techniques or modular cryptographic implementations that allow for flexible and dynamic adjustments based on network conditions [67]. Additionally, the development of cross-platform compatibility between PQC and existing TCPS is critical for a smooth transition.

### 6.4 Adoption by Vehicle Manufacturers and Stakeholders

Adopting PQC within the automotive industry poses both technical and economic challenges. Vehicle manufacturers and other TCPS stakeholders must assess the trade-offs between security and performance. PQC introduces computational overhead that can affect real-time decision-making systems in autonomous vehicles, potentially leading to degraded system performance. This challenge is particularly insistent in resource-constrained environments where efficient computation is critical [57]. Moreover, stakeholder adoption will require a shift in the current security standards and regulatory requirements. Collaboration between government agencies, automotive manufacturers, and technology providers is essential for establishing standardized protocols that ensure PQC's secure and efficient implementation into TCPS. Incentives, such as government-led regulations or performance benchmarks, may accelerate this transition. As more quantum-resistant algorithms become standardized, vehicle manufacturers will need to invest in hardware upgrades that support the additional computational demands of PQC [61, 66].

In summary, while PQC offers promising solutions for ensuring long-term sustainable security in TCPS, several challenges remain. Addressing the evolving side-channel threats, managing performance constraints, ensuring compatibility and interoperability, and encouraging adoption by industry stakeholders will be critical to fully realizing quantum-safe TCPS. Future research must continue to explore ways to balance security and performance, ensuring that PQC can be seamlessly integrated into real-world TCPS.

## 7 CONCLUSIONS

Quantum computers pose a significant threat to the cryptographic foundations of TCPS to meet its confidentiality and integrity requirements. Our research highlights the critical need to adopt PQC to mitigate these emerging risks. Through a comprehensive review of the NIST PQC Standardization guidelines and analysis of various PQC schemes, we identified promising public-key encryption/KEM candidates, such as CRYSTALS-Kyber, Classic McEliece, BIKE, and HQC, and digital signature candidates, such as CRYSTALS-Dilithium, Falcon, and SPHINCS+, which offer robust security claims against quantum attacks.

The case study presented in this paper on electronic toll collection further emphasized the vulnerabilities inherent in traditional cryptographic schemes in TCPS, showcasing how quantum threats could exploit these weaknesses. By employing the Microsoft Threat Modeling Tool, we identified potential risks and demonstrated how PQC can effectively mitigate them through the case study. This highlights the urgency of integrating PQC into existing and future TCPS data communication to ensure the integrity and confidentiality of critical TCPS.

Implementing PQC in real-world transportation systems involves various practical considerations, requiring a careful balance between security, performance, and resource constraints. Each PQC family offers distinct advantages and trade-offs, making their suitability dependent on specific TCPS requirements. For example, lattice-based schemes, such as CRYSTALS-Kyber and CRYSTALS-Dilithium, provide robust security and efficiency but may demand higher computational resources. Code-based schemes, such as Classic McEliece, excel in security with larger key sizes, making them suitable for bandwidth-insensitive applications, though less ideal for real-time safety-critical TCPS applications due to their high communication overhead. Multivariate polynomial-based schemes show potential for compact key sizes but are still undergoing active security evaluation, while hash-based schemes, attractive for constrained environments such as low-power or low-memory devices, require significant memory for key generation and storage. Isogeny-based schemes, offering high-performance potential, remain under continued security analysis.

Our experimental assessment of CRYSTALS-Kyber demonstrated its practical feasibility across various P2P communication scenarios, including wired (i.e., Ethernet) and wireless (i.e., LTE/C-V2X) environments for TCPS applications. The results indicate that Kyber's hierarchical security levels, while introducing increased computational and communication overhead with increasing levels, remain well-suited for different TCPS applications when paired with appropriate network infrastructures and communication-related performance requirements. For instance, high-bandwidth and low-latency Ethernet networks effectively mitigate Kyber's computational and communication overhead, making it a strong candidate for securing critical operations between static nodes. Examples include communication between a Payment Administration Center and DMV in electronic toll collection systems, as well as centralized traffic management systems, where secure and reliable data exchange between control centers and roadside infrastructure is critical. Similarly, results from the simulated LTE network, representing peer-to-peer C-V2X communication, reveal that while Kyber demonstrates feasibility in wireless communication scenarios, such as vehicle-to-vehicle data exchange, its current performance does not fully meet the stringent requirements of safety-critical TCPS applications. For example, the need to communicate 10 BSMs per second (i.e., 100 ms latency) imposes strict latency and throughput constraints, posing challenges for the wireless application of Kyber in its existing form. This highlights the necessity for further optimization, particularly in high-speed, low-latency wireless networks such as 5G, to ensure compliance with these demanding standards.

While the adoption of PQC represents a significant step forward in TCPS data and user security, challenges remain in ensuring interoperability, managing performance trade-offs, and addressing implementation constraints in the real-world transportation systems. Future research must prioritize the development of lightweight and hybrid PQC solutions tailored for real-time and resource-constrained environments, alongside extensive security evaluations under realistic TCPS conditions. Collaborative efforts among academia, industry, and government are essential to streamline standardization, incentivize adoption, and prepare transportation infrastructures for a quantum-secure future. Looking ahead, transitioning to PQC is essential to ensure the long-term sustainability and security of TCPS in the quantum era. By adopting these advancements, the transportation industry can safeguard critical infrastructures, fostering user trust and acceptance while ensuring confidentiality and the seamless operation of TCPS against emerging cyber threats.

**ACKNOWLEDGMENTS**

This work is supported by the National Center for Transportation Cybersecurity and Resiliency (TraCR) (a U.S. Department of Transportation National University Transportation Center) headquartered at Clemson University, Clemson, South Carolina, USA. Any opinions, findings, conclusions, and recommendations expressed in this material are those of the author(s) and do not necessarily reflect the views of TraCR, and the U.S. Government assumes no liability for the contents or use thereof. We also acknowledge Dr. Shuhong Gao, Professor of the Department of Mathematical Sciences at Clemson University, for reviewing this paper.